                                                                  \documentclass[12pt,preprint]{aastex}

\usepackage{CJK}

\shorttitle{The formation of  an inverse S-shaped active-region filament}
\shortauthors{Yan et al.}

\begin{document}
\begin{CJK*}{UTF8}{gbsn}

\title{The eruption of a small-scale emerging flux rope as the driver of an M-class flare and a coronal mass ejection}
\author{X.L. Yan\altaffilmark{1,4,8}, C.W. Jiang\altaffilmark{2}, Z.K. Xue\altaffilmark{1, 8}, J.C. Wang\altaffilmark{1, 7}, E.R. Priest\altaffilmark{3}, L.H. Yang\altaffilmark{1, 8}, D.F. Kong\altaffilmark{1, 4, 8}, W.D. Cao\altaffilmark{5}, H.S. Ji\altaffilmark{6}}

\altaffiltext{1}{Yunnan Observatories, Chinese Academy of Sciences, 396 Yangfangwang, Guandu Districk, Kunming 650216, Yunnan, P. R. China.(email: yanxl@ynao.ac.cn).}
\altaffiltext{2}{HIT institute of Space Science and Applied Technology, Shenzhen 518055, P. R. China.}
\altaffiltext{3}{Mathematics Institute, University of St Andrews, St Andrews, KY16 9SS, UK.}
\altaffiltext{4}{Key Laboratory of Solar Activity, National Astronomical Observatories, Chinese Academy of Sciences, Beijing 100012, P. R. China.}
\altaffiltext{5}{Big Bear Solar Observatory, 40386 North Shore Lane, Big Bear City, CA 92314, USA.}
\altaffiltext{6}{Key Laboratory for Dark Matter and Space Science, Purple Mountain Observatory, Chinese Academy of Sciences, Nanjing 210008, Jiangsu, P. R. China.}
\altaffiltext{7}{Graduate School of University of Chinese Academy of Sciences, Yuquan Road, Shijing- shang Block Beijing 100049, P. R. China.}
\altaffiltext{8}{Center for Astronomical Mega-Science, Chinese Academy of Sciences, 20A Datun Road, Chaoyang District, Beijing, 100012, P. R. China.}

\begin{abstract}
Solar flares and coronal mass ejections (CMEs) are the most powerful explosions in the Sun. They are major sources of potentially destructive space weather conditions. However, the possible causes of their initiation remain controversial. By using high resolution data observed by NST of BBSO, supplemented by Solar Dynamics Observatory (SDO) observations, we present unusual observations of a small-scale emerging flux rope near a large sunspot, whose eruption produced an M-class flare and a coronal mass ejection. The presence of the small-scale flux rope was indicated by static nonlinear force free field (NLFFF) extrapolation as well as data-driven MHD modeling of the dynamic evolution of the coronal 3D magnetic field. During the emergence of the flux rope, rotation of satellite sunspots at the footpoints of the flux rope was observed. Meanwhile, the Lorentz force, magnetic energy,  vertical current, and transverse fields were increasing during this phase. The free energy from the magnetic flux emergence and twisting magnetic fields is sufficient to power the M-class flare. These observations for the first time present the complete process from the emergence of the small-scale flux rope to the production of solar eruptions.

\end{abstract}


\section{Introduction}
Solar flares and coronal mass ejections (CMEs) are explosive phenomena observed in the solar atmosphere (Wang et al. 2002). A huge amount of free magnetic energy stored in the solar atmosphere is released during their eruptions (Forbes 2000; Priest \& Forbes 2002; Schmieder et al. 2015). This energy is transformed into radiative energy, bulk kinetic energy, thermal and non-thermal energy (Lin et al. 2015; Deng et al. 2013; Xiang \& Qu 2016). 

Previous observations demonstrate that the accumulation of free magnetic energy in the corona may be due to shearing motion(Wang et al. 1994; Moore et al. 2012), sunspot rotation (R$\acute{e}$gnier \& Amari 2006; Yan \& Qu 2007; Zhang et al. 2007; Yan et al. 2009; Sturrock et al. 2016; Zheng et al. 2017), and magnetic emergence or cancellation (Wang et al. 1993; Chen et al. 2000; Sterling et al. 2010) in the photosphere. However, the detailed processes for solar eruptions have been unclear for many years due to the difficulty of deducing the three-dimensional magnetic structure in the corona. Several possible models have been proposed for eruptions of flux ropes and the associated reconnection, namely, magnetic breakout, tether-cutting, magnetic non-equilibrium, kink instability and torus instability (Antiochos et al. 1999; Shibata 1995; Lin \& Forbes 2000; Amari et al. 2003; Carmichael 1964; Sturrock 1966; Hirayama 1974; Kopp \& Pneuman 1976; Moore et al. 2001; T{\"o}r{\"o}k et al. 2004; Kliem 2006; Schrijver 2008; Shen et al. 2012; Kliem et al. 2014; Chen et al. 2016; Shen et al. 2017). 

With the improvement of the observation, more and more fine structures of the Sun are revealed during solar eruptions. The model of flux rope in the solar eruptions has been proposed for many years (van Ballegooijen \& Martens 1989; Rust \& Kumar 1994, 1996). Until to recent years, the existence of flux ropes in the solar atmosphere is evidenced by using high resolution observational data (Tian et al. 2010; Cheng \& Ding 2016; Hou et al. 2016; Li et al. 2016; Zhou et al. 2016; Li et al. 2016; Lim et al. 2016). Zhang, Cheng, \& Ding (2012) reported that a flux rope with a twisted and writhed sigmoidal structure observed by SDO manifests itself as a hot channel before and during the solar eruption. The similar observation of magnetic cloud as the counterpart of a hot channel was studied by Song et al. (2015). They found that the hot channel is corresponding to a flux rope in solar corona. The existence of flux ropes can also be tracked out by material from a surge in a failed filament eruption (Yang et al. 2012). By using New Solar Telescope (NST) observation, Wang et al. (2015)  also found that a set of loops developed into a flux rope before a two-ribbon flare. Filippov et al. (2015) presented several filaments with highly twisted magnetic structure observed by TRACE and BBSO. These observations reveal that flux ropes may be ubiquitous in the solar atmosphere (Zhang et al. 2015).

However, it is still an open question how flux ropes form. Some models proposed that the photospheric activities can result in the formation of flux ropes via sunspot rotation, shearing motion, convergence flow, magnetic reconnection, and magnetic cancellation (van Ballegooijen \& Martens 1989; Fan 2009; Moore et al. 2001; Priest 2017). Recently, more and more observational evidences of flux rope formation are obtained from high resolution observation. Yan et al. (2012, 2015), Yang et al. (2015), and James et al. (2017) found that the formation of the S-shaped filaments or flux ropes is closely related to rotating sunspots that one foot of the filaments rooted in. This scenario is confirmed by the research of Vemareddy et al. (2016). They found that the magnetic connections of the sigmoid are driven by the slow motion of sunspot rotation, which finally transforms to a highly twisted flux rope structure. Joshi et al. (2014) presented a clear case that the formation of a compound flux rope via the merging of two nearby filament channels. Song et al. (2014) presented a direct observation of the flux rope formation process from a loop arcade during the eruption, which was associated with an M-class flare and a CME occurred at the southwest limb on 2013 November 21. Yan et al. (2016) reported that the formation of an inverse S-shaped flux rope is due to the reconnection of two groups of chromospheric fibrils from the 1 m New Vacuum Solar Telescope (NVST) observation (Liu et al. 2014). Kumar et al.( 2017) found that the reconnection between the cool H$\alpha$ loops in the chromosphere during the flares can form an unstable flux rope by using high resolution observation from the 1.6 m New Solar Telescope (NST) at BBSO. Some indirect evidence of the existence of flux rope can be deduced by using non-linear force free field (NLFFF) extrapolation (Cheng et al. 2010; Jiang et al. 2014; 2016a, b). 

The model of the emergence of flux rope has also been proposed before solar eruptions by some researchers (Rust \& Kumar 1994; Fan 2009). But the observational evidence is very rare. Up to now, only one case was reported on the emergence of flux rope by Okamoto et al. (2008). They found that a helical flux rope was emerging from below the photosphere into the corona along the PIL under the pre-existing prominence according to the observation of that the orientations of the horizontal magnetic fields along the PIL on the photosphere gradually changed with time from a normal-polarity configuration to an inverse-polarity one and the horizontal magnetic field region has blueshifted. However, Vargas Dominguez et al. (2012) argued that the signatures presented by Okamoto et al. (2008) are not sufficient indicators of twisted flux tube emergence. They found a decrease in the unsigned flux at the polarity inversion line (PIL) rather than the expected increase in the case studied by Okamoto et al. (2008). Furthermore, no shear motion and converging flows are detected in Okamoto's research. There are several simulations that were carried out to address the emergence of convection zone flux tubes. For instance, Archontis \& Hood (2013) presented the three-dimensional MHD simulations of the formation of jets triggered by the emergence and eruption of solar magnetic fields. The similar triggering process of jet and surge were obtained by the simulations of Moreno-Insertis \& Galsgaard (2013) and N{\'o}brega-Siverio, Moreno-Insertis, \& Mart{\'{\i}}nez-Sykora, (2016). Leake et al. (2014) performed three-dimensional magnetohydrodynamic simulations of the emergence of flux tubes from the convection zone into a pre-existing dipole coronal field. They found the external reconnection between the emergence flux tube and the overlying magnetic field lines is vital to the eruption process. That is to say, this external reconnection helps the further expansion of the coronal flux rope into the corona. MacTaggart \& Haynes (2014) considered the formation of two flux ropes in a magnetohydrodynamic solar flux emergence simulation. In their simulation, a shearing motion along the polarity inversion line (PIL) is found during the formation of two flux ropes. Moreover, when the first rope approaches the top boundary of the domain, it is dissipated. The second flux rope emerges due to the decrease of the overlying coronal magnetic fields by reconnection between the emerging flux region and the ambient magnetic fields.

Here we use NST and SDO muti-wavelength data to investigate the process from the emergence of a small-scale flux rope to the production of an M-class flare and a small coronal mass ejection.  Observations and methods are presented in Section 2. The results are shown in Section 3. Conclusions and discussions are given in Section 4.

\section{Observations and methods}
\subsection{Observations}
TiO (7057 \AA) images and H-alpha images are obtained from the Visible Imaging Spectrometer (VIS) on the 1.6 m New Solar Telescope at BBSO (Cao et al. 2010; Goode et al. 2010). TiO images have a spatial resolution of 0.$^\prime$$^\prime$0342  per pixel and the time cadence is 15 s. H-alpha images have a pixel size is 0.$^\prime$$^\prime$0295 and the cadence is 43 s for VIS to complete a scan at a 0.2 \AA\ step from -1.0 \AA\ to + 1.0 \AA\ around the H-alpha line center. H-alpha blue-wing (-0.6 \AA), H-alpha center, and H-alpha red-wing (+0.6 \AA) images are used in this study.

Full-disk UV and EUV images with a 12-sec cadence and a spatial resolution of 0.$^\prime$$^\prime$6  per pixel observed by SDO/AIA (Lemen et al. 2012) are employed to show the process of flux rope eruption. Vector magnetograms from Space Weather HMI Active Region Patch (SHARP) series observed by the Helioseismic and Magnetic Imager (HMI) (Schou et al. 2012; Bobra et al. 2014; Centeno et al. 2014) have a pixel scale of about 0.$^\prime$$^\prime$5 and a cadence of 12 minutes. They are derived using the Very Fast Inversion of the Stokes Vector algorithm (Borrero et al. 2011). The minimum energy method (Metcalf 1994; Metcalf et al. 2006; Leka et al. 2009) is used to resolve the 180 degree azimuthal ambiguity. The images are remapped using Lambert (cylindrical equal area) projection centered on the midpoint of the AR, which is tracked at the Carrington rotation rate (Sun 2013). 
\subsection{Methods}
\subsubsection{Current calculation} 
The calculation of current is based on Ampere's law:
\begin{equation}\label{equation1}
{\bf J}=\frac{c}{4\pi}(\bigtriangledown \times {\bf B})
\end{equation}
in which $\mu_0$ is the vacuum magnetic permeability and ${\bf B}$ denotes the vector magnetic fields. The current density perpendicular to the solar surface can be calculated from the Ampere's law by using HMI vector magnetograms according to the equation
\begin{equation}\label{equation2}
j_z=\frac{c}{4\pi}(\bigtriangledown \times {\bf B})_z=\frac{c}{4\pi}(\frac{\partial B_x}{\partial y}-\frac{\partial B_y}{\partial x})
\end{equation}
where $B_x$ and $B_y$ are the two components of the  photospheric horizontal magnetic field.
The distribution of $j_z$ on the solar surface is obtained every 12 minutes. Note that, all the equations in this paper are written in cgs.

\subsubsection{Calculation of magnetic flux} 
The integrated positive ($\rm\phi_{zp}$) and negative ($\rm\phi_{zn}$) magnetic fluxes are calculated from
\begin{equation}\label{equation4}
\rm{\phi_{zp}}=\it\int B_{z+}dA
 \end{equation}
 \begin{equation}\label{equation5}
\rm{\phi_{zn}}=\it\int |B_{z-}|dA
 \end{equation}
where $B_{z+}$ and $B_{z-}$ are the vertical positive and negative magnetic fields and $dA$ denotes the area differential.\\

\subsubsection{NLFFF extrapolation} 
The vector magnetic fields obtained by SDO/HMI are taken as the boundary condition to extrapolate the magnetic fields from the photosphere to the corona. An optimization algorithm proposed by Wheatland et al. (2000) and implemented by Wiegelmann (2004) is used to extrapolate the three-dimensional NLFFF structure of the active region. A preprocessing procedure (fff\_temp\_pre.pro in SSW) is used to deal with the bottom boundary vector data before the extrapolation. This removes most of the net force and torque that would result in an inconsistency between the forced photospheric magnetic field and the force-free assumption in the NLFFF models (Wiegelmann et al. 2006).\\

\subsubsection{Lorentz force calculation} 
Lorentz force at the photospheric surface  is divided into radial and horizontal components, which can be written as follows:
\begin{equation}\label{equation4}
\rm{F_{r}}={1\over{8\pi}}\it\int_{A_{ph}} {(B_{h}}^{2}-{B_{r}}^{2})dA
\end{equation}
 \begin{equation}\label{equation5}
 \rm{F_{h}}={1\over{8\pi}}\it\int_{A_{ph}} B_{h}B_{r}dA
 \end{equation}
Here, $B_{h}$ and $B_{r}$ represent the horizontal  and radial components of B, which are parallel and vertical to the photosphere, respectively. $A_{ph}$ is the area of the white box in the photosphere. These results are derived from the integral of the divergence of the Maxwell stress tensor.

The total Lorentz force acting on plasma at and above the solar photosphere can be written in terms of surface integrals of products of components of the photosphere field {\bf B}=$(B_{x}, B_{y}, B_{y})$ by assuming the contributions from the top and side boundaries are negligible (Fisher et al. 2012). Thus, the horizontal $(F_{x}, F_{y})$ and vertical $(F_{z})$ components can be written as follows: 
\begin{equation}\label{equation5}
\rm{F_{x}}={1\over{4\pi}}\it\int_{A_{ph}} B_{x}B_{z}dA  
\end{equation}
 \begin{equation}\label{equation6}
 \rm{F_{y}}={1\over{4\pi}}\it\int_{A_{ph}} B_{y}B_{z}dA
 \end{equation}
 \begin{equation}\label{equation7}
 \rm{F_{z}}={1\over{8\pi}}\it\int_{A_{ph}} ({B_{x}}^{2}+{B_{y}}^{2}-{B_{z}}^{2})dA
 \end{equation}

\subsubsection{Free energy calculation} 
We use NLFFF extrapolation to obtain the magnetic fields above the photosphere. Then the extrapolated NLFFF fields ($B_{ext}$) and potential fields ($B_{pot}$) are used to calculate the free energy from the equation
\begin{equation}\label{equation8}
\rm{E_{free}}={1\over{8\pi}}\it\int_{v} ({B_{ext}}^{2}-{B_{pot}}^{2})dV  
\end{equation}
where $B_{ext}$ and $B_{pot}$ indicate the extrapolated NLFFF fields and potential fields, respectively. V is the cube with a height of ten arcseconds whose bottom is marked by the box in Fig. 3b. 

\subsubsection{Data-driven simulation} 
The vector magnetograms from Space Weather HMI Active Region Patch (SHARP) series with a pixel scale of about 0.$^\prime$$^\prime$5 and a cadence of 12 minutes were used to derive the magnetohydrodynamic (MHD) model. The images were remapped using Lambert (cylindrical equal area) projection centered on the midpoint of the AR, which is tracked at the Carrington rotation rate (Sun 2013). The full set of time-dependent 3D MHD equations is solved by using the bottom boundary condition driven continuously by the changing photospheric vector magnetic fields from SDO observations (Jiang et al. 2016).  Here, the background plasma is initially in a hydrostatic, isothermal state with $T=10^{6}$~K (sound speed $c_{S}=128$~km~s$^{-1}$) in solar gravity. Its density is configured to make the plasma $\beta$ as small as $2\times 10^{-3}$ (the maximum Alfv{\'e}n $v_{\rm A}$ is $4$~Mm~s$^{-1}$) to mimic the coronal low-$\beta$ and highly tenuous conditions. The plasma thermodynamics are simplified to be adiabatic since we focus on the evolution of the coronal magnetic field. The bottom boundary of the model is assumed to be the coronal base, and also the magnetic field measured on the photosphere is used as a reasonable approximation to the field at the coronal base.

\section{Results}
There was a large active region in the south hemisphere, named NOAA 12403, on 2015 August 24 (see Fig. 1a). The active region was located  near the center of the solar disk with a $\beta$$\gamma$$\delta$ field configuration of the sunspot group. The SDO observations lasted from the emergence of small satellite sunspots ``S1" , ``S2", and ``S4" near the big sunspot with negative polarity to the occurrence of the flux rope eruption in this active region (see the positions of the satellite sunspots in Figs. 2a-2d). Moreover, Big Bear Solar Observatory (BBSO) high-resolution observations detected the details of the eruption. 

The evolution of the vector magnetic fields, NLFFF extrapolations, electric current, and Lorentz force during the formation and eruption of the flux rope is presented in Fig. 2. The field of view of Fig. 2 is marked by a yellow box in Fig. 1b. The emergence process of sunspots ``S1" , ``S2", and ``S4" is clearly seen in supplementary movie 1. The sunspot ``S2" first emerges at the right of the sunspot ``S3" (see Fig. 2a) and then it moves from the south to the north with a strong shearing motion between S2 and S3 (see Figs. 2b-2c and supplementary movie 1).  The white and black patches indicate the positive and negative magnetic polarities in Figs. 2a-2h. The blue arrows in Figs. 2a-2d indicate the transverse magnetic fields. The transverse magnetic fields around the four sunspots exhibit clockwise vortex structure. The change of the integrated magnetic flux in the white box of Fig. 2b is shown in Fig. 3a. The NLFFF extrapolations in Figs. 2e-2g present the emergence of the structure of a flux rope, which includes the twisted magnetic field lines and a highly sheared arcade. Note that the Space-weather SDO/HMI Active Region Patches (SHARP) vector magnetogram data were used to extrapolate the 3D magnetic field of this active region. After the eruption of the flux rope, the twisted structure disappears and the connections of the magnetic field lines also change (see Fig. 2h). The configuration of magnetic fields become more potential. 

Figs. 2i-2l show evolution of the electric current during the emergence of the satellite sunspots and after the eruption of the flux rope. An increase of the current of sunspot S3 is associated with its emergence. The red and blue patches indicate the positive and negative electric current, respectively. The green and black contours outline the positive and negative magnetic polarities. Note that the levels of the contours are $\pm$500 G,  $\pm$1300 G, $\pm$2100 G. Since the sunspots ``S2" and ``S3" exhibit a significant rotation during their evolution (see supplementary movie 2), the direction of the Lorentz force also exhibits vortex shape (see Figs. 2m-2p). The rotation direction of the sunspots is consistent with the directions of the Lorentz force. During the emergence of the sunspot S2, the upward and transverse Lorentz force between the sunspots S2 and S3 increases rapidly from 07:36 UT to 11:12 UT.

Figure 3 presents the change of magnetic flux, vertical currents, Lorentz force, free energy, GOES X-ray flux, and transverse fields in the white box of Fig. 2b. The black and blue lines indicate the change of positive and negative magnetic flux in Fig. 3a.  It is significant that the negative magnetic flux increases during the emergence of sunspot ``S2", while the positive magnetic flux almost remains constant. The change of negative vertical currents is similar to that of the negative magnetic flux (see Fig. 3b).  The Lorentz force has a rapid increase at 08:00 UT (the beginning of the emergence of the sunspot S3) and at about 14:20 UT (three hours before onset of the flux rope eruption). The transverse fields and the Lorentz force almost reach their maximum values simultaneously (see Figs. 3c and 3e). The presence of a positive vertical Lorentz force causes the flux rope to rise. Previous investigation indicates that Lorentz force plays an important role in the solar eruptions (Bi et al. 2016). The transverse fields and the Lorentz force begin to decrease simultaneously at 15:20 UT. Because the transverse fields and Lorentz force are calculated in the photosphere, the decrease of these values suggests that the emergence of the field has now finished. The free magnetic energy begins to  increase when the flux rope appears (see Fig. 3d). After the emergence of the flux rope has finished, the decrease of the free energy may be due to the eruption of some small chromospheric fibrils before the flux rope eruption.

Figure 4 shows the photospheric evolution of small satellite sunspots near the large sunspot observed by BBSO and SDO/HMI. The TiO image and line-of-sight magnetogram observed by BBSO/NST and SDO/HMI are presented in Figs. 4a and 4b, respectively. Four small satellite sunspots are labelled by ``S1", ``S2", ``S3", and ``S4" in Figs. 4a and 4b. All of the sunspots have positive polarity, except for ``S2" which has negative polarity. Meanwhile, the four sunspots exhibit a clockwise rotation (see supplementary movie 1).  To obtain the rotation angle of sunspot ``S3", the circumference of a circle with a radius of 2$^\prime$$^\prime$ is mapped by a polar coordinate r-$\theta$ frame centered on the middle of S3, as illustrated in Figure 4c. The resolution used in the angular direction is $1^{\circ}$ to make the time slices (Zheng et al. 2016). The variation of the rotation angle with time is calculated by tracing the penumbra of sunspot ``S3" (see Fig. 4f). The green lines that are tracing two penumbral features show the rotation of penumbral filaments around the center of the circle. The rotation angles are $60^{\circ}$ after 42 minutes and $110^{\circ}$ after 200 minutes. The rotation of the sunspots implies that the twist is transferred from below the photosphere to the upper atmosphere (Sturrock \& Hood 2016). This is inferred from the increase of the twist of the flux rope associated with the rotation of the sunspot. Figs. 4c-4e demonstrate the evolution of these sunspots before and after the M-class flare. The M-class flare starts at 17:40 UT, peaks at 17:46 UT, and ends at 17:49 UT (see the red curve in Fig. 3d).  The penumbra between sunspot ``S2" and sunspot ``S3" gradually decreases before the onset of the M-class flare (see the blue arrows in Figs. 4c and 4d). The disappearance of the penumbra is associated with the occurrence of the M-class flare (see Fig. 4e). That is to say, the disappearance of the penumbra between sunspots S2 and S3 is due to the rising of the flux rope. 

The entire eruptive process of the small flux rope was caught by the observation of BBSO/NST. Figure 5 shows the eruption process for the M-class flare observed by BBSO/NST (see supplementary movie 3). The three columns from left to right are the H-alpha blue-wing (-0.6 \AA), H-alpha center, and H-alpha red-wing (+0.6 \AA) images. The contours superimposed on the first row images are the levels of the line-of-sight magnetic fields. The green and blue contours indicate the positive and negative magnetic polarity. The levels of the contours are $\pm$ 300 G and $\pm$ 800 G. The red arrows mark the positions of the sunspots. The yellow arrows in Figs. 5a-5c indicate the flux rope before its eruption. It is clear that the two footpoints of the flux rope are mainly rooted in sunspots ``S2" and ``S3". Some threads of the flux rope also can be seen to root in the outer penumbra of the large sunspot, which is implied by the NLFFF extrapolations (see Figs. 2e, 2f, and 2g). Before the flux rope eruption, a series of small chromospheric fibrils at the upper end of the flux rope first begin to erupt from 16:37:19 UT to 17:00:37 UT (see the green arrows in Figs. 5d-5i).  At 17:09:05 UT, the upper part of the flux rope becomes darker and starts to erupt. A strong blue shift can be seen in the H-alpha blue-wing images. The closed magnetic field lines are opened and followed by a strong red shift in the H-alpha red-wing images. The flux rope starts to erupt at 17:40:12 UT associated with a strong red shift at the upper end of the flux rope. At the beginning of the flux rope eruption, the whole body of the flux rope rotated counterclockwise. After the eruption of the flux rope, the closed magnetic field lines are opened again (see the H-alpha blue-wing and the H-alpha center observation at about 17:50:40 UT). These observations clearly show that, before the onset of the large eruption, there are small eruptions of chromospheric fibrils as a precursor. Using Swedish 1 m Solar Telescope (SST) and Hinode observational data, Guglielmino et al. (2010) found that reconnection between the small-scale emerging flux region and the overlying coronal field produces the brightenings in the chromosphere, transition region, and corona, as well as chromospheric surges. Our results are very similar to these observations. The change in topology in our study is also similar to what is found in flux emergence simulations when the emerging field pushes into an overlying field and reconnects with it (Galsgaard et al. 2005; MacTaggart \& Haynes 2014; Leake et al. 2014). The precursor of flux rope eruption is also similar with the simulation results of MacTaggart et al. (2015), which show how the surges occur when a small-scale emerging flux region reconnects with ambient fields.

During the flux rope eruption, the plasma is ejected into the upper atmosphere (see Figs.6a-6f).  The blue arrows in Figs. 6a-6f indicate the flux rope during its eruption in 304 \AA\ and 171 \AA\ images. At 18:24 UT, LASCO C2 observed a coronal mass ejection (see Figs. 6d-6f).

A data-driven MHD model for solar active-region evolution (Jiang et al. 2016a) is employed to provide further understanding of the process of emergence and eruption of the field, which is shown in Fig. 7. Here the MHD simulation starts from 08:00 UT on August 24, when the small sunspot S2 just emerges near the small sunspot S3. The initial conditions consist of a NLFFF extrapolated from the vector magnetogram at this start time and a highly tenuous plasma in a hydrostatic state. Subsequently, the model is continuously supplied with a changing bottom boundary condition from the data stream of photospheric vector magnetograms observed by SDO/HMI. To self-consistently input the vector magnetograms at the boundary, the method of projected characteristics based on the wave-decomposition principle of the full MHD system (Wu et al. 2006) was used. The result of the simulation can be seen in Figs. 7a-c and supplementary movie 4. Meanwhile, the evolution of electric current in a cross-section along (see Figs. 7d, 7e, and 7f) and perpendicular (see Figs. 7g, 7h, and 7i) to the flux rope is calculated from the simulation. It is clear that the magnetic field emerges in the course of the simulation from t=0 to 41, as driven by the evolution of the photospheric field.  During the emergence of the field, the current increases rapidly along the flux rope and the rising of the flux rope can also be seen from the cross-section of the current perpendicular to the axis of the flux rope. At around t=55, both the twisted magnetic flux and the corresponding current density decreased quickly, indicating the eruption of the flux rope which releases some of the non-potential energy of the pre-eruption field, after which the magnetic structure is closer to potential. These observation and simulation confirm the existence of flux rope in the solar eruptions (Xue et al. 2016).

\section{Conclusions and discussions}
Observational evidences of the emergence of a small-scale flux rope as the trigger of the M-class flare and a coronal mass ejection are present by using high spacial and temporal data from NST and SDO. The small sunspot with negative polarity emerges near the small sunspot with positive polarity. During the  emergence of the small-scale flux rope, the negative flux, electric current, the Lorentz force, and transverse magnetic fields increase gradually. The magnetic free energy calculated from the extrapolated 3D magnetic fields is found to be enough to power the M-class flare. The NLFFF extrapolation and the data-driven MHD simulation also reveal the emergence of the flux rope. Moreover, the sunspots where the flux rope rooted in exhibit clockwise rotation. These observations present a clear picture of solar eruptions: emergence of the small-scale flux rope leads to the flare and the CME.

There are several evidences that can confirm the emergence of the flux rope. First, the small sunspot with negative polarity emerged near the small sunspot with positive polarity. The negative magnetic fluxes and the electric current obviously increase at the emergence of the flux rope. Most of magnetic field lines of the flux rope extrapolated from the NLFFF rooted in the two small sunspots. Second, the transverse magnetic fields increase at first and decrease about two hours before its eruption. Third, the change of upward Lorentz force also increases at the beginning of the flux rope and then decrease before the eruption of the flux rope, which is similar to that of the transverse of the magnetic fields. Fourth, the sunspot rotation was observed during the emergence of the flux rope. Fifth, the twisted magnetic structures can be obtained from the NLFFF extrapolation. Sixth, the data-driven MHD modeling of the dynamic evolution of the coronal 3D magnetic field recreates the emergence of the flux rope. Seventh, the free magnetic energy increases rapidly at the stage of the emergence of the flux rope. We observed sunspot rotation, shearing between the two opposite polarities, and the separation of the two small sunspots during emergence of the field. This scenario is very consistent with simulations of the emergence of a twisted flux tube from the sub-photosphere to the corona (Magara 2006; Fan 2009; Hood et al. 2009; Hood, Archontis, \& MacTaggart 2012; Archontis, Hood, \& Tsinganos 2014). Several surge-like eruptions were observed as the precursor of the eruption of the small-scale twisted flux rope. This agrees well with the prediction from emerging flux simulations of Galsgaard et al. (2005) and MacTaggart et al. (2015). Furthermore, as the flux rope emerges, the Lorentz force produces shearing along the polarity inversion line between the two small sunspots and the flux rope is carried upwards by the positive Lorentz force. These observations match very well with the simulation of MacTaggart \& Hood (2009).

Helical flux ropes are fundamental magnetic structures in some eruptive-flare models (Priest et al. 1989; van Ballegooijen \& Martens 1989; Rust \& Kumar 1994; Amari et al. 2003; Gibson et al. 2006; Fan 2009). However, it is very difficult to determine whether the flux ropes form prior to or during eruptions. Many observational evidences support that flux ropes already exist before the onset of solar eruptions (Liu et al. 2003; Yan et al. 2012; Chintzoglou et al. 2015; Su et al. 2015; Vemareddy et al. 2016; Cheng \& Ding 2016; James et al. 2017).  Others suggest that flux ropes form during solar eruptions (Song et al. 2014; Cheng et al. 2010). Recent observations show that unwinding motion is often found during active-region filament eruptions (Yan et al. 2014a, b; Zhang et al. 2015), which implies that the active-region filaments may have twisted magnetic structures. Srivastava et al. (2010) presented a direct observation evidence for the existence of a twisted magnetic structure in the corona. Due to the spatial and temporal resolution of the data, where the twist of flux ropes comes from is not easy to observe. Sunspot rotation is one candidate way for injecting magnetic twists into the corona and the twisted magnetic structures or solar filaments are indeed observed after sunspot rotation (Yan et al. 2012, 2015; James et al. 2017). There are some cases that  flux ropes are formed by magnetic reconnection of sheared arcade loops (Chen et al. 2014; Yang et al. 2016; Wang et al. 2017; Xue et al. 2017). Another way is emergence of twisted flux ropes from below the photosphere in view of many simulations (Magara 2006; Fan 2009; Hood et al. 2009; MacTaggart et al. 2009; Hood, Archontis, \& MacTaggart 2012).

Usually, the formation process of small-scale flux ropes cannot be observed due to lack of high resolution photospheric magnetograms and chromospheric observations. Fortunately, the observations of NST and SDO caught a clear process of the emergence of a small flux rope. In this study, the photospheric and chromospheric observations of NST, supplemented by SDO observations, can present the whole process of the formation of a flux rope associated with the emergence of a small sunspot (pore). Moreover, the two small sunspots at the footpoints of the flux rope exhibited a clockwise rotation. Sunspot S2 rotates around its center and moves away from sunspot S3 during emergence of the flux rope. This scenario is consistent with the emergence of a twisted flux rope. The increase of twist in the flux rope with time follows the rotation of small sunspots. It implies that sunspot rotation may play an important role in the increase of the twist of the flux rope, which propagates from below the photosphere to the upper atmosphere. In the past, many simulations have included high twist to generate flux ropes and CMEs in the atmosphere. Our results tie in well with such simulations (Magara 2006; Fan 2009; Hood et al. 2009; MacTaggart \& Hood 2009; Hood, Archontis, \& MacTaggart 2012; Galsgaard et al. 2005; MacTaggart et al. 2015). Our observations give very strong evidence for small-scale highly-twisted ropes emerging to form CMEs as the simulations predict.
.

The eruptive mechanism for solar eruptions is also an open issue. There are several models to address this question. Ideal MHD instability is one of these models (Hood \& Priest 1979). T{\"o}r{\"o}k et al. (2003) simulated the evolutionary process of a twisted flux rope and found that when the critical value of twist is larger than 2.75$\pi$, the flux rope cannot keep stable and will erupt. Some observations can also confirm their results (Srivastava et al. 2010; Kumar et al. 2012; Yan et al. 2014a,b). In this study, the twist of the flux rope extrapolated from NLFFF is about 3 turns. When the small-scale magnetic region emerged from under the penumbra of the large sunspot, the overlying magnetic field was perpendicular to the axis of the flux rope. But the flux rope exhibited a counterclockwise rotation at the onset of its eruption, the axis of the flux rope became antiparallel to the overlying magnetic field. External reconnection between the emerging field and the overlying magnetic field may weaken the tension of the coronal field. This scenario is very similar to the simulations of MacTaggart \& Hood (2009) and Leake et al. (2014). Therefore, reconnection between the emerging field and the pre-existing extended field of the large sunspot also plays an important role in flux rope eruption.

\acknowledgments 
We would like to thank the NST, SDO/AIA, SDO/HMI teams for high-cadence data support. This work is sponsored by the National Science Foundation of China (NSFC) under the grant numbers 11373066, 11603071, 11503080, 11633008, 11533008, Key Laboratory of Solar Activity of CAS under numbers KLSA201603, KLSA201508, Yunnan Science Foundation of China under number 2013FB086, CAS ``Light of West China" Program. Youth Innovation Promotion Association CAS (No.2011056), and the national basic research program of China(973 program, 2011CB811400). The BBSO operation is supported by NJIT, US NSF AGS-1250818, and NASA NNX13AG14G grants, and the NST operation is partly supported by the Korea Astronomy and Space Science Institute and Seoul National University and by the strategic priority research program of CAS with Grant No. XDB09000000.

\begin{figure}
\epsscale{.90}
\plotone{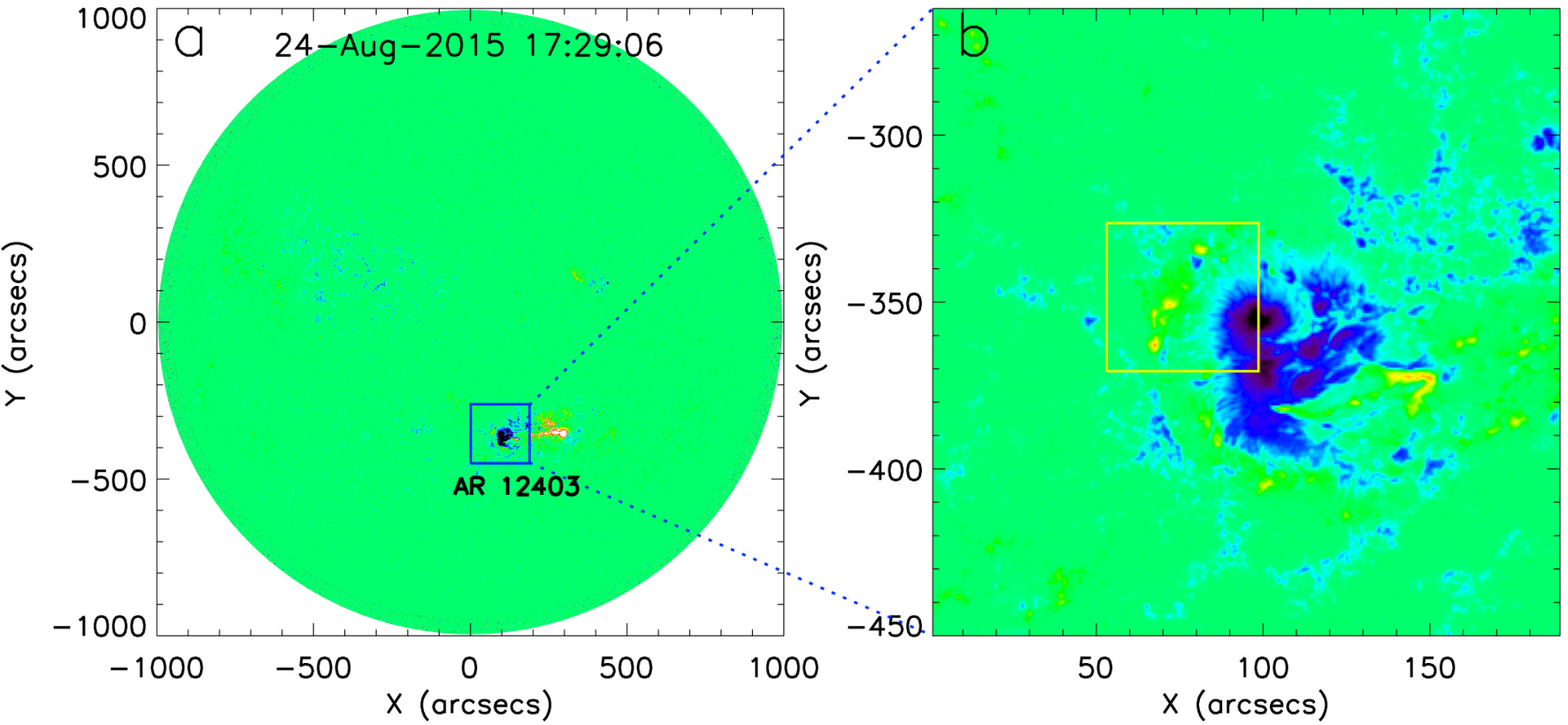}
\caption{Left panel: Full-disk line-of-sight magnetogram showing active region NOAA 12403 on the solar disk. Right panel: the following sunspots of the active region. The yellow box outlines the field of view of Fig. 2 and Fig. 5. \label{fig1}}
\end{figure}

\begin{figure}
\epsscale{.90}
\plotone{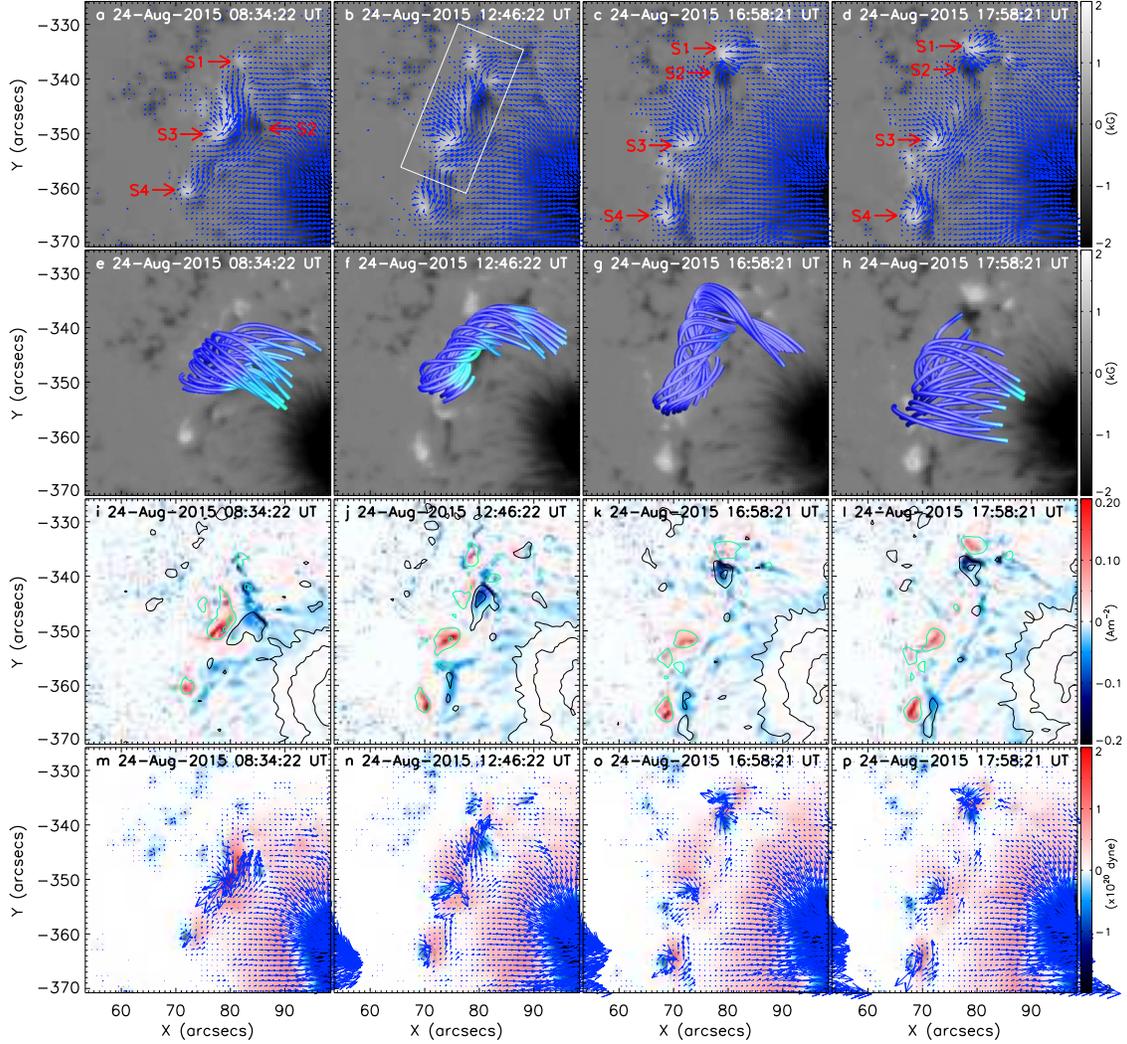}
\caption{Evolution of the vector magnetograms, the magnetic structure of the NLFFF extrapolation, the electric current, and Lorentz force. (a-d) Vector magnetograms observed by SDO/HMI. The red arrows indicate the small satellite sunspots. The white box shows the region used to calculate the magnetic flux, free energy, current, and Lorentz force for this event. (e-h) The magnetic structure of NLFFF extrapolations. (i-l) The evolution of the current. The red and blue patches show the positive and negative current. The black lines outline the sunspots. (m-p) The evolution of the Lorentz force. The red and blue patches show the upward and downward Lorentz force. The blue arrows show the directions of the transverse Lorentz force. \label{fig1}}
\end{figure}

\begin{figure}
\epsscale{.90}
\plotone{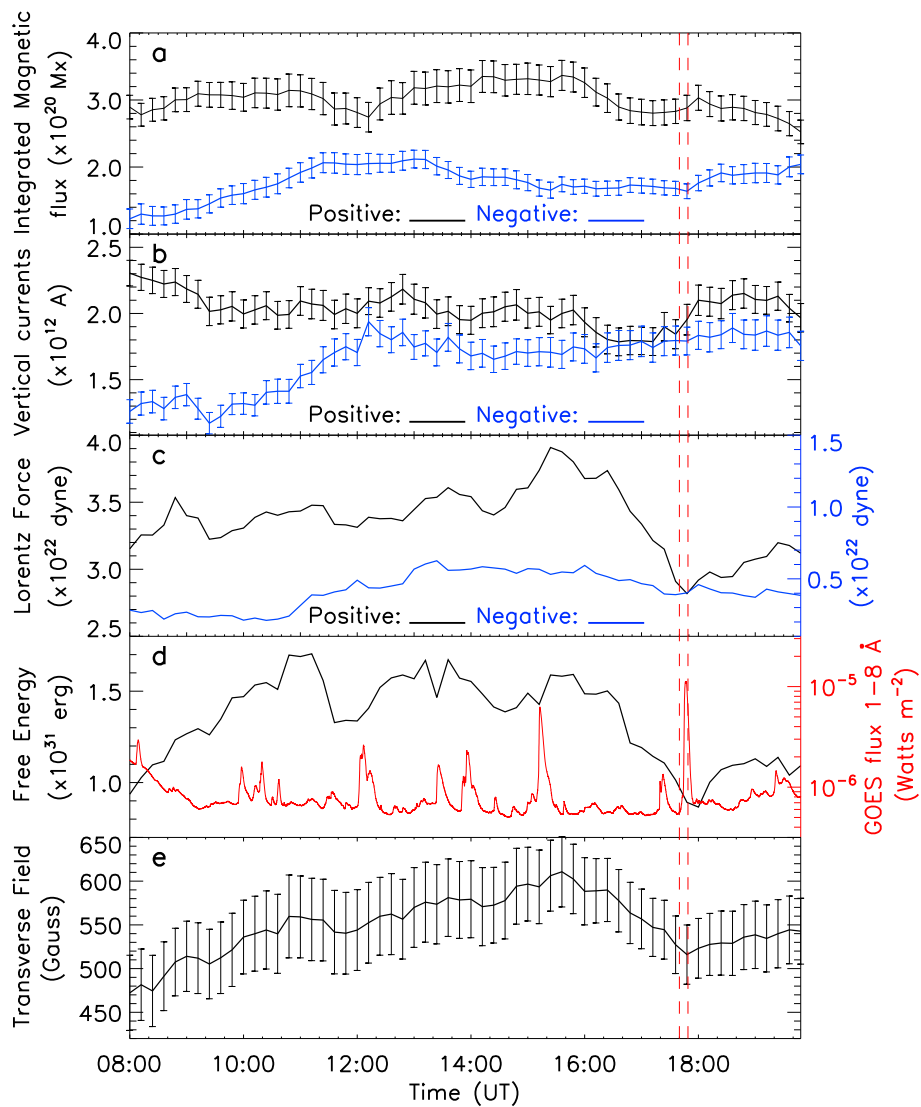}
\caption{Evolution of the magnetic flux(a), vertical current (b), Lorentz force(c),  free energy (d), and the transverse magnetic fields (e). The red curve shows the GOES soft X-ray flux in panel d. The vertical red dashed lines show the onset and ending time of the M-class flare. \label{fig1}}
\end{figure}

\begin{figure}
\centering
\includegraphics[angle=0,scale=1.0]{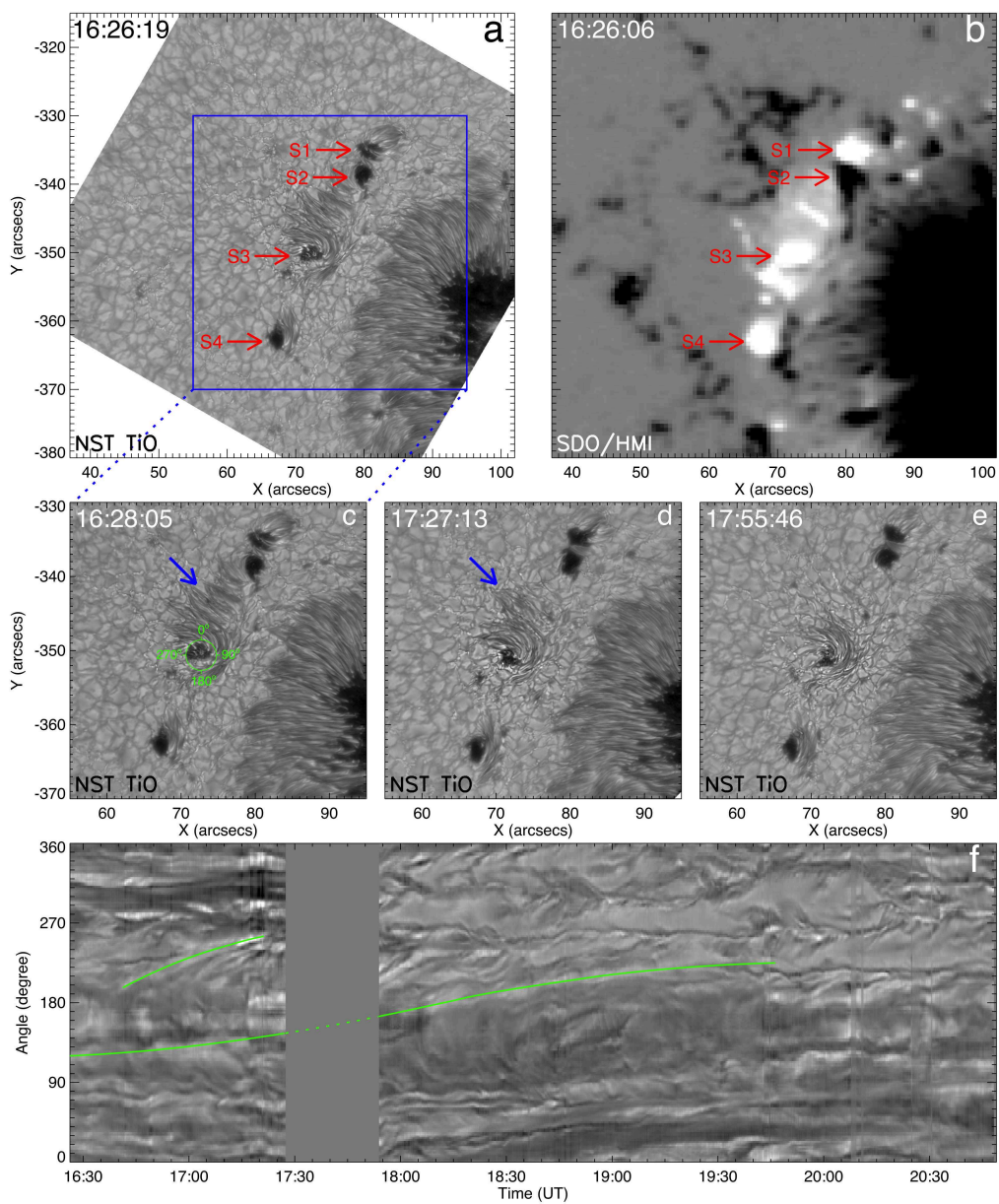}\\
\caption{Evolution of small satellite sunspots in active region NOAA 12403. (a) The NST TiO image showing the appearance of the small satellite sunspots indicated by the red arrows. (b) The corresponding HMI magnetogram. (c-e) The change of the small satellite sunspots in TiO images. The change of the penumbra between S2 and S3 is indicated by the blue arrows. The green circle is the position of the time slice. (f) Time-slice plot acquired along the circle marked by a green circle in Fig. 1c, and the green curved lines indicate the rotation of the penumbra filament around the center of the sunspot S3.}
\end{figure}

\begin{figure}
\epsscale{.80}
\plotone{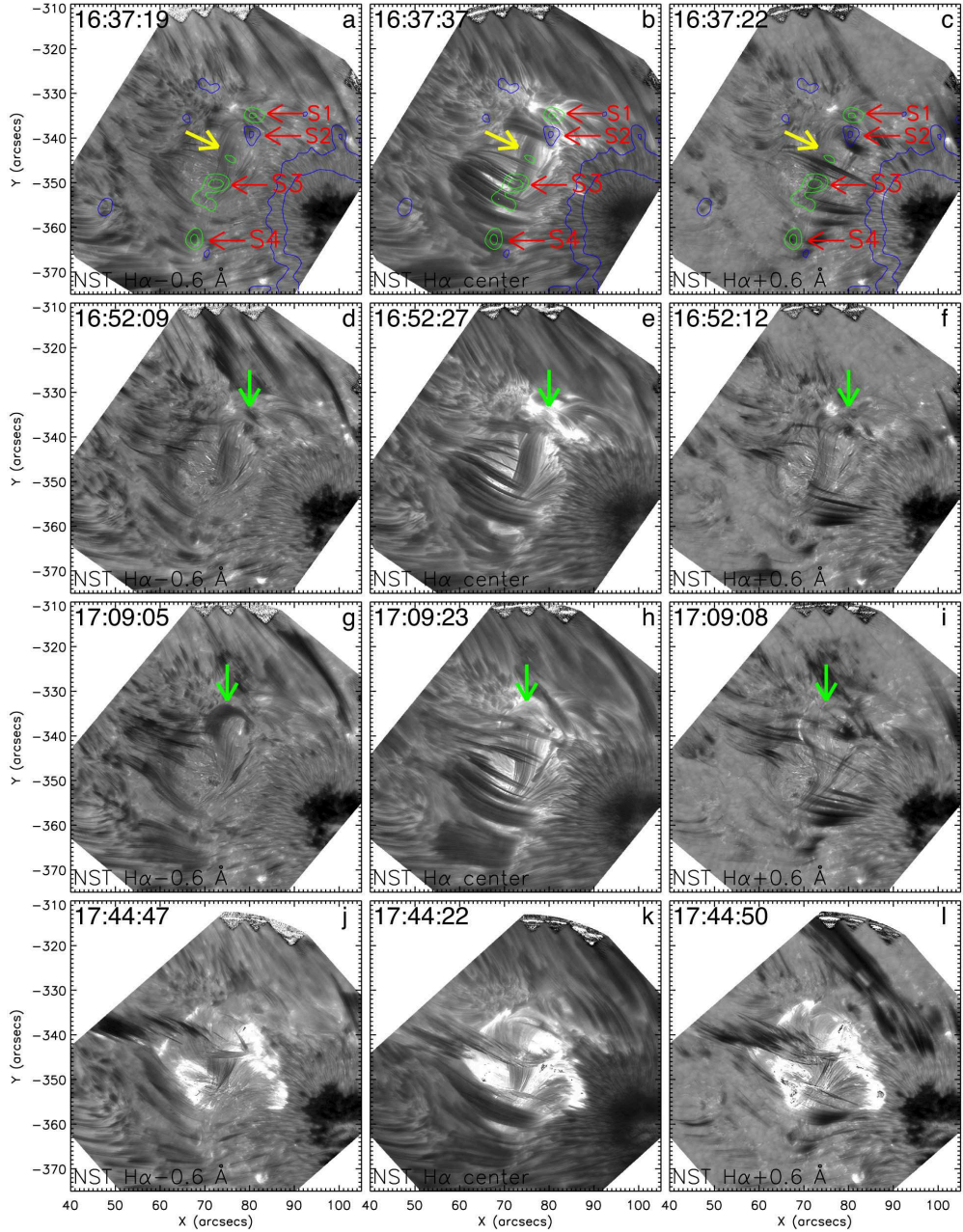}
\caption{Eruption process of the M-class flare. (a, d, g, j): the NST H-alpha blue-wing images acquired at -0.6 \AA .  (b, e, h, k): the NST H-alpha center images.  (c, f, i, l): the NST H-alpha red-wing images acquired at +0.6 \AA. The first row images are overlaid by green (blue) contours representing positive (negative) polarity. The yellow arrows indicate the flux rope and the red arrows indicate the small satellite sunspots. The green arrows indicate the small threads at the upper part of the flux rope, which erupt before the flux rope eruption. \label{fig1}}
\end{figure}%

\begin{figure}
\epsscale{.80}
\plotone{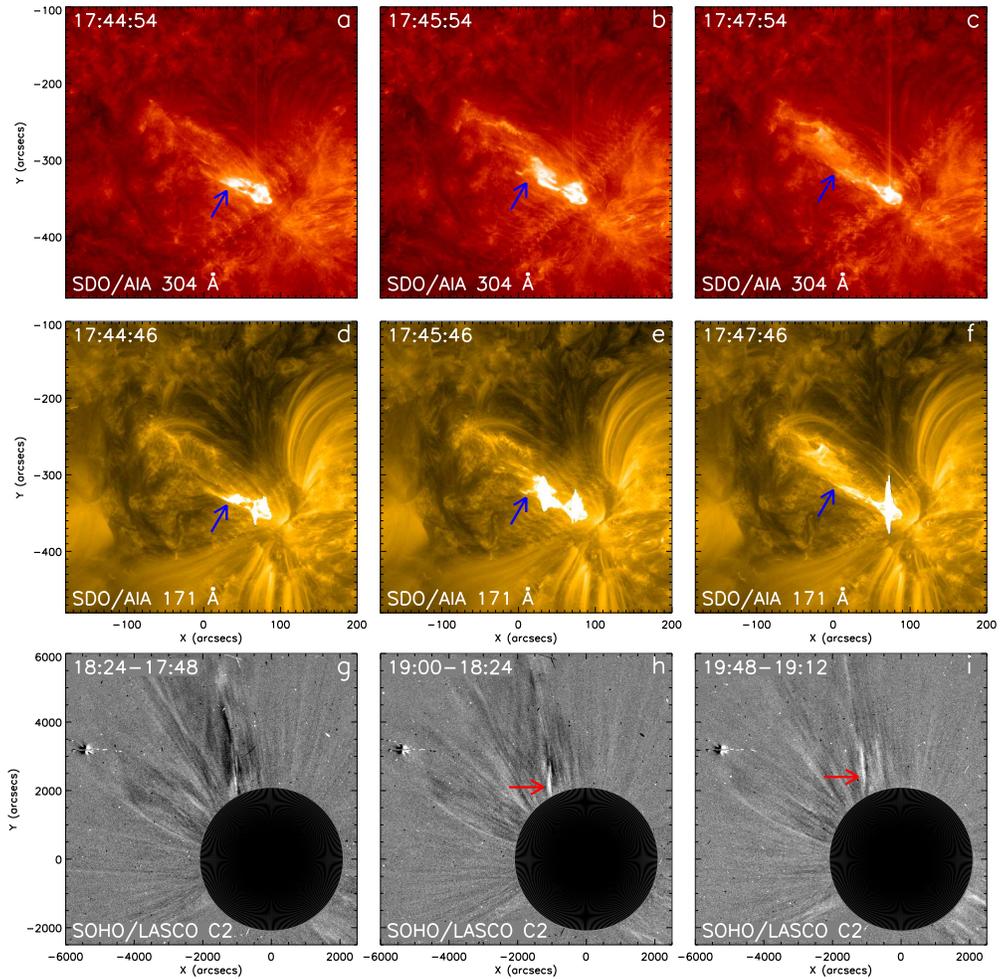}
\caption{The M-class flare and the coronal mass ejection observed in SDO 304 \AA\ and 171 \AA\ images, and SOHO/LASCO C2. The blue arrows indicate the plasma ejection from the active region while the red arrows indicate the small coronal mass ejection after the flux rope eruption. \label{fig1}}
\end{figure}

\begin{figure}[ht]
{\centering
 \includegraphics[width=12cm]{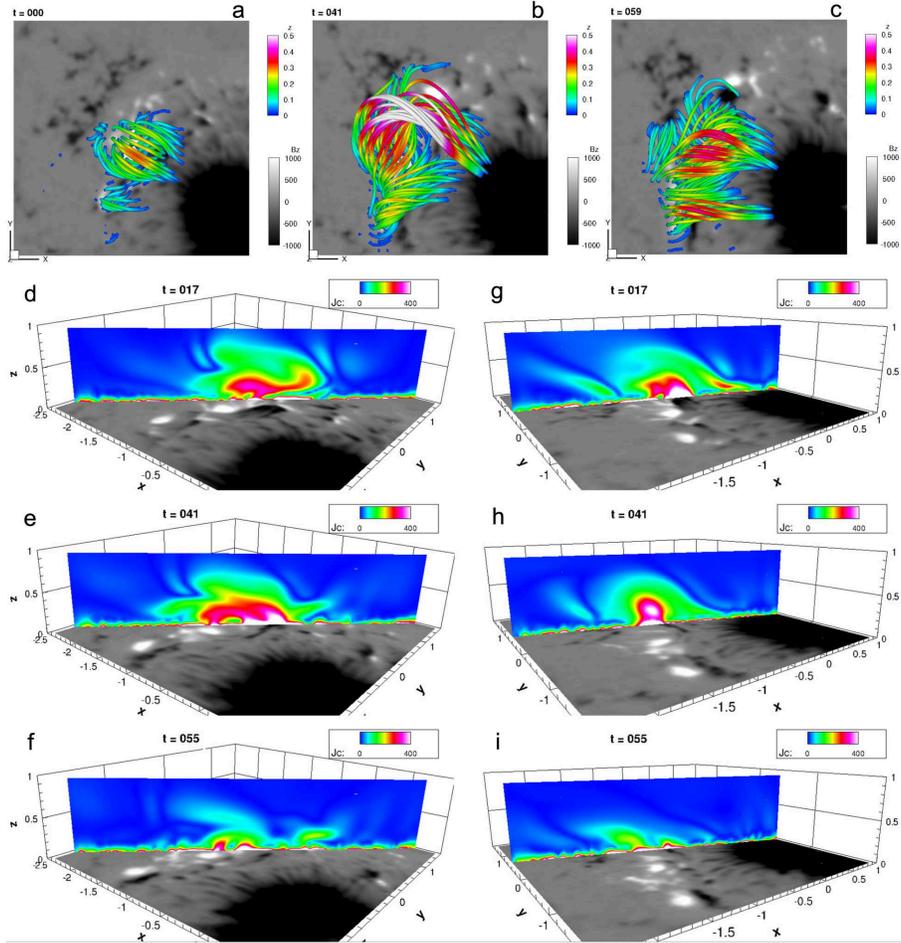}
\caption{Data-driven MHD simulation of flux rope emergence and eruption.  a-c: The emergence of the flux rope. d-f: the current along the flux rope. g-i: the current perpendicular to the flux rope. }
\label{fig6}}
\end{figure}

\end{CJK*}
\end{document}